\newif\ifpdflatex    
\pdflatextrue           

\documentclass[fleqn,usenatbib,useAMS,twocolumn,numberedappendix]{aastex}
\usepackage{amsmath,esint}
\usepackage{url,aasmacros}
\usepackage{natbib}
\usepackage{soul, CJK}
\usepackage{multirow}
\usepackage{tablefootnote}
\usepackage{appendix}
\usepackage{bm}
\usepackage{xspace}
\usepackage{color}
\usepackage{enumitem}
\usepackage{booktabs}

\usepackage{graphicx}	
\usepackage{amsmath}	
\usepackage{amssymb}	
\usepackage{bm}		
\usepackage{url}
\usepackage{amsbsy}
\usepackage{xspace}
\usepackage{hyperref}
\usepackage{longtable}
\usepackage[colorinlistoftodos]{todonotes}
\usepackage[flushleft]{threeparttable}

\usepackage[T1]{fontenc}
\usepackage{ae,aecompl}

\def\lesssim{\mathrel{\hbox{\rlap{\hbox{\lower5pt\hbox{$\sim$}}}\hbox{$<$}}}}
\def\gtrsim{\mathrel{\hbox{\rlap{\hbox{\lower5pt\hbox{$\sim$}}}\hbox{$>$}}}}

\newcommand{\angstrom}{\textup{\AA}\xspace}

\usepackage{upgreek}                                                  
\usepackage{xspace}                                                       
\newcommand{\um}{$\upmu$m\xspace}            
 
\def\apjl{ApJ}%
\def\apjs{ApJS}%
\def\aap{A\&A}%
\shorttitle{SPHEREx transients}
\shortauthors{Karambelkar et al.}

\begin{document}
\title{SPHEREx as a frontier for infrared transients: Classification of new Galactic FU Ori outbursts and classical novae}

\author[0000-0003-2758-159X]{Viraj Karambelkar}
\email{vk2588@columbia.edu}
\altaffiliation{NASA Hubble Fellow}
\affiliation{Columbia University, 538 West 120th Street 704, MC 5255, New York, NY 10027}

\author[0000-0002-8989-0542]{Kishalay De}
\affiliation{Columbia University, 538 West 120th Street 704, MC 5255, New York, NY 10027}
\affiliation{Center for Computational Astrophysics, Flatiron Research Institute, 162, 5th Ave, New York, NY 10010}

\author{Lynne A. Hillenbrand}
\affiliation{Cahill Center for Astrophysics, California Institute of Technology, Pasadena, CA 91125, USA}

\author{J. L. Sokoloski}
\affiliation{Columbia University, 538 West 120th Street 704, MC 5255, New York, NY 10027}

\author[0000-0002-7197-9004]{Danielle Frostig}
\affiliation{Center for Astrophysics | Harvard \& Smithsonian, 60 Garden Street, Cambridge, MA 02138, USA}

\author[0000-0002-1264-2006]{Julianne Dalcanton}
\affiliation{Center for Computational Astrophysics, Flatiron Research Institute, 162, 5th Ave, New York, NY 10010}
\affiliation{University of Washington, Department of Astronomy, Box 351580, Seattle WA 98103, USA}






\begin{abstract}
We demonstrate proof-of-concept of a new strategy for studying infrared (IR) transients enabled by the newly launched SPHEREx space mission, by leveraging its synergy with the NEOWISE space mission. With its fifteen year baseline and all-sky mid-IR coverage, NEOWISE provides an excellent avenue to discover thousands of slowly evolving infrared outbursts. With its all-sky spectro-photometric coverage and mid-IR sensitivity matching NEOWISE, SPHEREx is uniquely positioned to provide low-resolution IR spectra for the vast majority of these outbursts, several of which are too obscured for ground-based spectroscopic classification. As a demonstration of this approach, we present SPHEREx spectra for eight Galactic transients identified in NEOWISE. This sample includes two previously known FU Orionis-type (FUOr) outbursts  whose SPHEREx spectra exhibit clear signatures of cool molecular absorption and three known classical novae showing strong emission lines in SPHEREx. Using these sources as templates, we identify two new FUOrs and one previously missed Galactic nova. Our results highlight the potential of SPHEREx for systematic explorations of the relatively underexplored dynamic infrared sky. \\ \\
\end{abstract}

\section{Introduction}
Over the last decade, the study of transient and time variable astronomical phenomena has undergone a revolution, driven by wide-field automated time-domain surveys (e.g., \citealt{Bellm2019, Tonry2018, Chambers16}). However, the majority of such surveys have been conducted at optical wavelengths, and are thus inherently biased against the dustiest stellar eruptions that emit primarily in the infrared (IR).

The relatively high cost of IR detectors has historically limited systematic exploration of the dynamic IR sky. Previous IR transient searches have relied on targeted observations of selected galaxies or regions using facilities equipped with IR cameras (e.g., \citealt{Kasliwal2017ApJ, Minniti2010, Kool2018, Saito2024}), often without dedicated transient-identification pipelines. Only recently, dedicated surveys, such as the all-sky Palomar Gattini IR (PGIR, \citealt{De2020}), the Wide-field Infrared Transient Explorer (WINTER, \citealt{Lourie2020, Frostig2025_winter}), and the PRime Focus Infrared Microlensing Experiment (PRIME, \citealt{Konndo2023}) have begun to systematically monitor large areas of the sky in the near-IR. 

A recent advance in IR time-domain astronomy has come from novel data analysis techniques applied to the fifteen-year baseline of observations from the NEOWISE survey \citep{Mainzer2014}. These efforts have discovered thousands of time-varying mid-IR sources over the entire sky (e.g., \citealt{De2023Nature, Paz2024, Zihan2025}), many of which are so obscured that they do not have any optical counterparts \citep{Zuckerman2023, John2024, Frostig2026}. Ground-based near-IR spectroscopic follow-up of some of these sources revealed a diverse population of previously hidden dusty transients (e.g., \citealt{De2023Nature,De2024_m31}), highlighting the potential of NEOWISE to explore the obscured transient sky. A general challenge for characterizing these sources is that their highly obscured nature renders ground-based spectroscopic follow-up expensive or sometimes completely infeasible. 

The newly launched Spectro-Photometer for the History of the Universe, Epoch of Reionization and Ices Explorer (SPHEREx) space mission \citep{Crill2020, Alibay2023} provides an exciting opportunity to address this challenge. SPHEREx is a spectrophotometer employing linear variable filters that will effectively conduct a $0.7-5$\um~low-resolution (R$\sim40-130$) spectroscopic survey of the entire sky four times over its two year mission. While full SPHEREx spectra will be released only for a pre-defined target list \citep{Ashby2023} designed to achieve the mission's core science goals \citep{Dore2016, Dore2018} --- cosmology \citep{Dore2018AAS}, galaxy evolution \citep{Zemcov2018}, and the origin of ices \citep{Ashby2023} --- all calibrated SPHEREx images \citep{Akeson2025} will be publicly released throughout the survey. These data enable users to extract low-resolution spectra for sources of interest using publicly available tools \footnote{\href{https://irsa.ipac.caltech.edu/applications/spherex/tool-spectrophotometry}{https://irsa.ipac.caltech.edu/applications/spherex/tool-spectrophotometry}}. Notably, SPHEREx has sensitivities in the mid-IR (3--5\um) bands comparable to NEOWISE, with its highest spectral resolution in this wavelength range, offering the possibility of providing mid-IR spectra for a large fraction of NEOWISE transients. 

In this Letter, we present a proof-of-concept of this scheme. We first extract SPHEREx spectra for five NEOWISE transients with existing classifications and show that their key features are recovered. Using these features, we propose spectroscopic classifications for three previously unclassified transients. In Section\,\ref{sec:neowise_transients}, we describe our NEOWISE transient identification and selection, Section\,\ref{sec:spherex_spectra} describes the SPHEREx spectra, and Section\,\ref{sec:classifications} describes the transient identifications. We conclude with a summary of our results and the way forward in Section\,\ref{sec:summary}.

\section{NEOWISE mid-IR transients}
\label{sec:neowise_transients}
The WISE satellite \citep{Wright2010}, reinitiated as the NEOWISE mission \citep{Mainzer2014}, conducted a mid-infrared survey of the entire sky in the W1 (3.6 $\mu$m) and W2 (4.5 $\mu$m) filters with a cadence of $\approx$6 months between 2014 to 2024. As part of an ongoing program to search for mid-IR transient outbursts using NEOWISE data, we applied image-differencing techniques to time-resolved co-added images created as part of the unWISE project \citep{Lang2014, Meisner2019}. The full technical details will be presented in a forthcoming publication (De et al., in prep.). Briefly, the ZOGY algorithm \citep{zogy} was used to perform image subtraction on all coadded NEOWISE images relative to the full-depth co-added images from the WISE survey (taken in 2010--2011). All transients that were identified in the difference images with a detection threshold of S$_{\rm{corr}}>5$ were cataloged and assigned names of the form WTPxxyyyyyy, where where `xx' indicates the year of first detection and `yyyyyy' is a six-letter alphabetical code. This search has yielded several thousand distinct, time-varying, mid-IR sources, including dusty eruptive Galactic and extragalactic transients such as classical novae \citep{Zuckerman2023, Ramesh_2025}, outbursting FU Orionis-type (FUOr) young stellar objects \citep{Frostig2025}, dusty supernovae \citep{Mo2025}, stellar mergers \citep{Karambelkar2025}, and tidal disruption events \citep{Masterson2024} as well as exotic transients like a planetary engulfment event \citep{De2023Nature} and a failed supernova \citep{De2024_m31}.

For this pilot study designed to demonstrate the potential of SPHEREx, we focus on the brightest NEOWISE transients with SPHEREx coverage. We therefore restrict our sample to Galactic IR outbursts detected in the last two years (post-2023) of the NEOWISE survey. Specifically, we search for NEOWISE transients within |b|$<10$\,deg. that have W1 and W2 detections after 2023, and are either slowly evolving or are bright in the latest NEOWISE epoch (flux density $>10^{4}$\,$\mu$Jy). 

\begin{figure*}
    \centering
    \includegraphics[width=\linewidth]{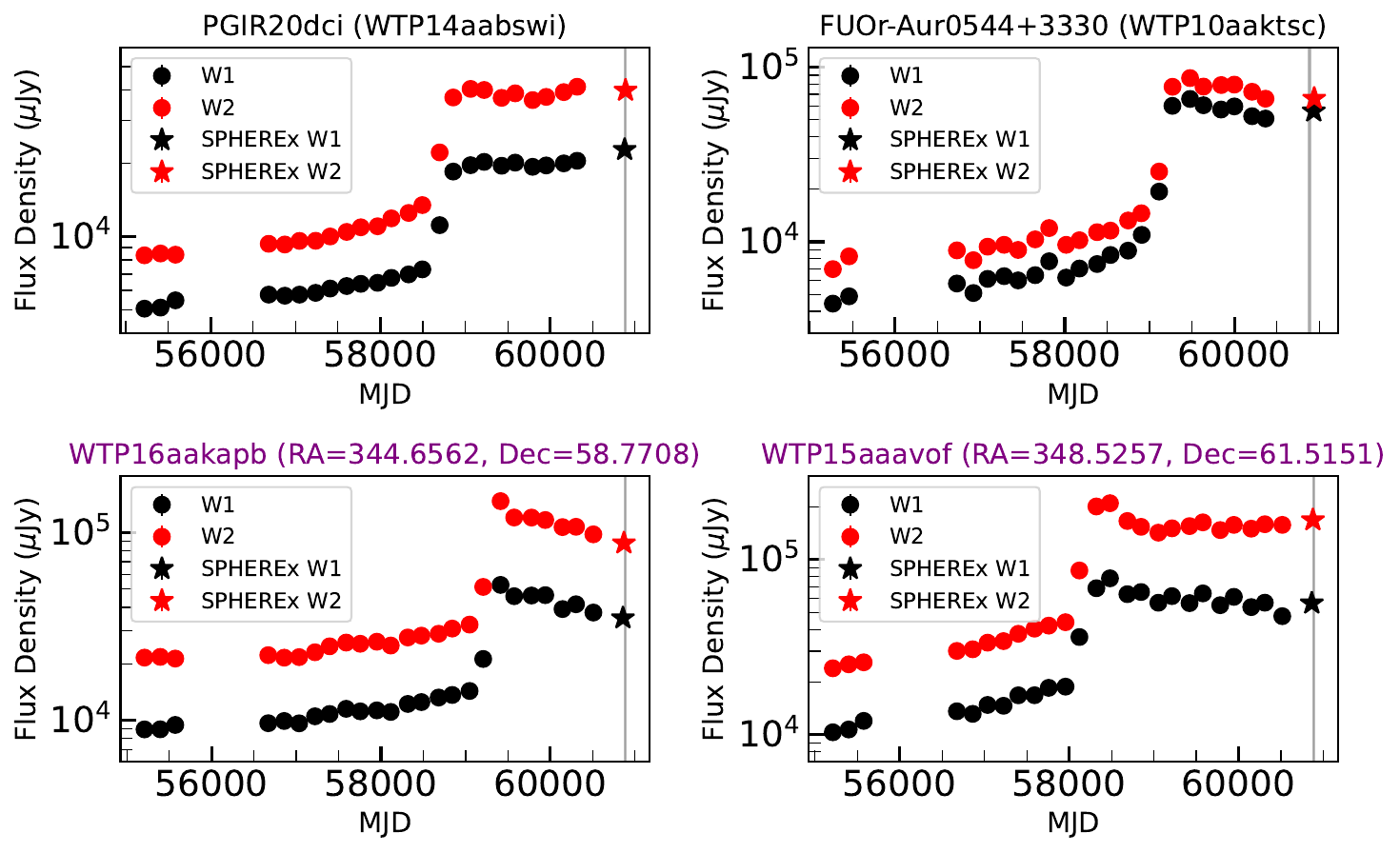}\\
    \includegraphics[width=\linewidth]{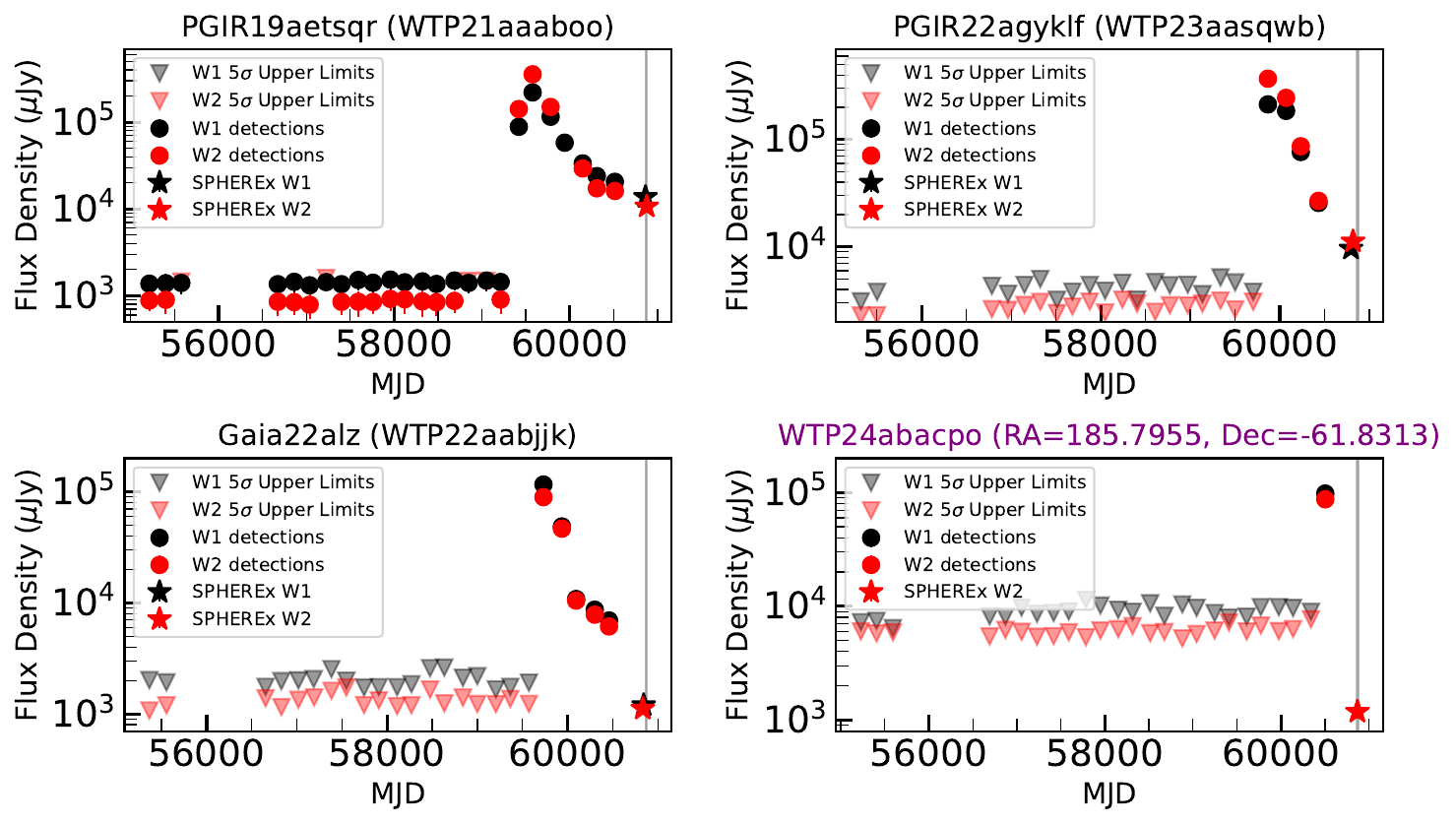}
    \caption{NEOWISE lightcurves of the eight transients presented in this paper. Circles mark the NEOWISE photometry, while stars indicate synthetic photometry derived from SPHEREx. The previously known transients are indicated with their published names, while the NEOWISE names of the three newly identified transients are marked in purple with their coordinates. The gray vertical shaded region marks the epochs of SPHEREx observations.}
    \label{fig:lcs}
\end{figure*}

Here, we present eight of the brightest transients selected from the sources satisfying these criteria. These include three previously known classical novae, two previously known FUOrs, and three new transients. Figure \ref{fig:lcs} shows the NEOWISE lightcurves of these transients. Future works will present a more comprehensive analysis of a broader sample of sources.

\section{SPHEREx spectrophotometry}
\label{sec:spherex_spectra}
SPHEREx surveys the sky in a mode that provides a low-resolution 0.75--5\um~spectrum of every pixel on the sky over a period of $\sim$two weeks \citep{Crill2020}\footnote{See \href{https://spherex.caltech.edu/page/survey}{https://spherex.caltech.edu/page/survey} for further details about the mission.}. The spectral resolution varies from R$\sim40$ in Band 1 (0.75--1.12\,\um) to R$\sim130$ in Band 6 (4.41--5.01\,\um), while detector pixel scale is 6.2" per pixel. These characteristics enable SPHEREx to provide infrared spectra for the majority of slowly evolving, bright IR transients identified in the final years of the NEOWISE survey. SPHEREx completed its first all-sky survey in December 2025, and Level~2 Calibrated images covering nearly the entire sky were publicly released about two months after data acquisition \citep{SPHEREx_IPAC_dataset}.

We use the online public SPHEREx Spectrophotometry Tool \footnote{\href{https://irsa.ipac.caltech.edu/applications/spherex/tool-spectrophotometry}{https://irsa.ipac.caltech.edu/applications/spherex/tool-spectrophotometry}} to obtain SPHEREx spectrophotometry for the transients selected above. The spectrophotometry tool performs forced point-spread function photometry on the Level 2 calibrated SPHEREx images \footnote{see \href{https://irsa.ipac.caltech.edu/data/SPHEREx/docs/SPHEREx_Expsupp_QR.pdf}{https://irsa.ipac.caltech.edu/data/SPHEREx/docs/\\SPHEREx\_Expsupp\_QR.pdf} for further details.} We filter the measurements to photometric points with signal-to-noise$>3$ and are below the saturation thresholds. We find that for most of our sources, 90\% of the SPHEREx spectrum is accumulated within a time span of 17 days or fewer, with the exception of WTP23aasqwb (33 days) and WTP15aaavof (41 days). As these sources are already at phases several years post-eruption, we do not expect substantial evolution over these timescales. We also note that the SPHEREx spectra of these bright sources generally have very few statistical outliers, and these can be often identified based on a relatively higher value of the \texttt{fit\_ql} metric compared to the remainder of the spectrum (see e.g. the spectrum of WTP16aakapb in Fig. \ref{fig:yso_spectra} which shows an outlier at 3.4\,\um~that is likely a bad photometric measurement, as it has \texttt{fit\_ql=25} while most of its spectrum has \texttt{fit\_ql}$<5$).

We use the SPHEREx spectra to estimate synthetic W1 and W2 band magnitudes for the eight transients and plot them in Figure \ref{fig:lcs} (stars). We find that the SPHEREx synthetic photometry is overall consistent with the brightness evolution trends seen in NEOWISE.

\section{Transient identifications}
\label{sec:classifications}
In this section, we first identify hallmark spectral features of the known transients using their SPHEREx spectra. We then use these diagnostics to propose classifications for the new transients.

\subsection{Transients with molecular absorption: Outbursting FUOr sources}
\begin{figure*}[hbt]
    \centering
    \includegraphics[width=\textwidth]{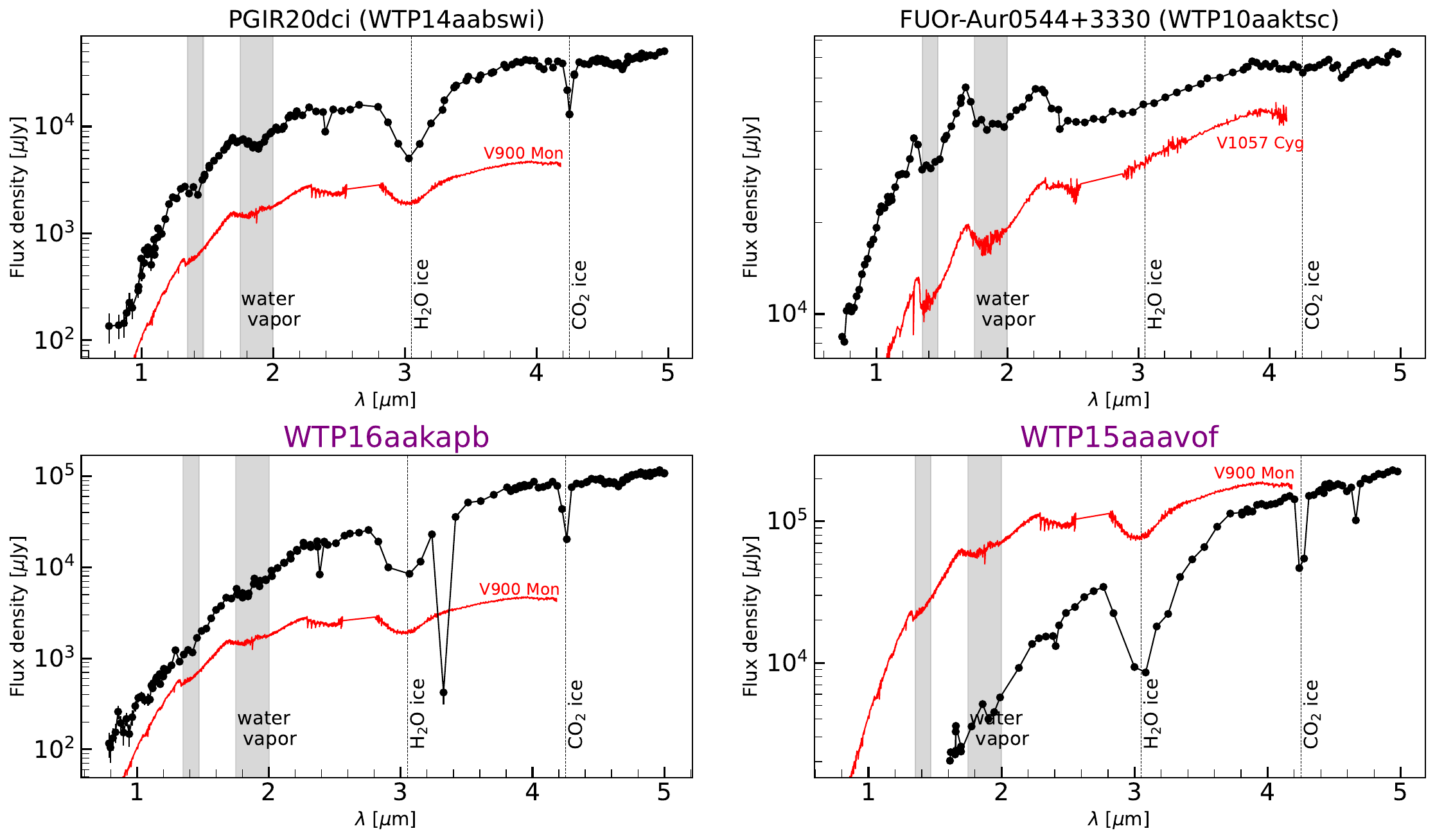}
    \caption{SPHEREx infrared spectra for the four transients showing molecular absorption features --- two previously known and two new FU Ori-type young stellar outbursts. Plotted in red, for comparison, are medium resolution, ground-based IR spectra of known FU Ori outbursts V900 Mon, V1057 Cyg \citep{Connelley2018}. For WTP16aakapb, the sharp drop in flux at $\approx3.4$\,\um~likely results from a bad photometric measurement (see Section \ref{sec:spherex_spectra}).}
    \label{fig:yso_spectra}
\end{figure*}

Two transients in our sample are previously known FUOr outbursts --- PGIR20dci (WTP14aabswi; \citealt{Hillenbrand2021}) and FUOr-Aur0544+3330 (WTP10aaktsc; \citealt{Hillenbrand2025}). Their NEOWISE light curves exhibit a characteristic two-step brightening \citep{Hillenbrand2021, Tran:2024}: a slow brightening for $\sim$3000 days, followed by a sharp ``ramp-up" of W1 and W2 fluxes over $\sim500$ days, followed by a plateau or a slow decline. The SPHEREx spectra of these sources are shown in the top row of Figure \ref{fig:yso_spectra}. At near-IR (1--2\um) wavelengths, both spectra show red continua with prominent water vapor absorption features. These features agree well with the NIR spectra for these sources presented in \citet{Hillenbrand2021, Hillenbrand2025}. No emission lines are detected at the low spectral resolution of SPHEREx. At mid-IR wavelengths (2-5\um), the spectrum of PGIR20dci shows strong absorption features due to H$_{2}$O ice centered at 3.05\um~and CO$_{2}$ ice at 4.25\um, features commonly seen in disks of FUOrs \citep{Connelley2018} and other young stellar objects \citep{Gibb2004,Boogert2004, Boogert2008,Oberg2008,Zasowski2009, Bottinelli2010, Aikawa2012}. Overall, the spectrum of PGIR20dci resembles that of the FUOr V900\,Mon, whose medium resolution spectrum from \citealt{Connelley2018} is shown for comparison in Figure\,\ref{fig:yso_spectra}. In contrast, FUOr-Aur0544+3330 does not show any ice absorption features, instead exhibiting a smoothly rising continuum, likely because it is less obscured than PGIR20dci. Its mid-IR spectrum is similar to that of the FUOr V1057\,Cyg \citep{Connelley2018} also shown in Figure \ref{fig:yso_spectra}.

\begin{figure}
    \centering
    \includegraphics[width=\linewidth]{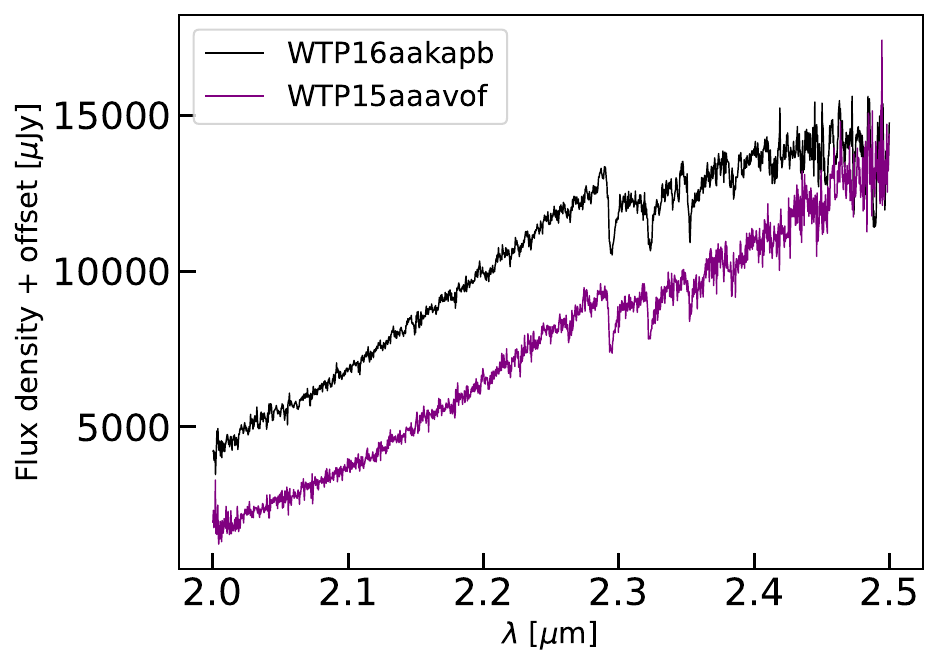}
    \caption{Follow-up medium resolution K-band spectra for the two new FUOrs identified in this paper, showing CO absorption bands that are not resolved in the low-resolution SPHEREx spectra.}
    \label{fig:yso_irtf}
\end{figure}
Two new transients in our sample --- WTP16aakapb (RA=344.6562, Dec=58.7708) and WTP15aaavof (RA = 348.5257, Dec=61.5151) --- have NEOWISE light curves and SPHEREx spectra that closely resemble the FUOrs described above. Their NEOWISE light curves (Figure \ref{fig:lcs}), show the two step brightening, with flux increases of order $\sim$10x over the last fourteen years. Their SPHEREx spectra (Figure \ref{fig:yso_spectra}) resemble PGIR20dci, with red continua and strong absorption features of H$_{2}$O and CO$_{2}$ ice. WTP16aakapb also shows weak water-vapor absorption bands in the 1.8--2\um~range, while WTP15aaavof is extremely obscured and has no SPHEREx detections shortward of 2\,\um. Applying a standard extinction law to the SPHEREx spectra and comparing them with the NIR spectrum of V900\,Mon (A$_{V}\approx13$\,mag), we estimate total  A$_{V}\approx29$ and $\approx40$\,mag for WTP16aakapb and WTP15aaavof respectively. These high values are consistent with high optical depths of water-ice absorption $\tau\approx1.5$ for WTP16aakapb and $\approx1.6$ for WTP15aaavof, derived following \citealt{Connelley2018}, which suggest A$_{V}\gtrsim30$\,mag \citep{Connelley2018}. Both sources are located near active star-forming regions in the Perseus arm. WTP16aakapb is located in the star-forming compact HII region Sh2-152, while WTP15aaavof is located in the star-forming region NGC7538 and has been previously identified as a Class I YSO \citep{Chavarria2014, Sharma2017}. Adopting distances to these regions of $\approx2.4$\,kpc for Sh2-152 \citep{Russell2007} and $\approx2.7$\,kpc for NGC7538 \citep{Moscadelli_2009}, the reddening values derived above, and the peak W2 flux densities, we estimate peak $\nu$L$_{\nu} \approx50$\,L$_{\odot}$ and $\approx130$\,L$_{\odot}$ for WTP16aakapb and WTP15aaavof respectively. These values are consistent with other FUOrs. 

As follow-up observations, we obtained medium resolution near-IR spectra for both these sources using the SPeX spectrograph \citep{Rayner2003} on the NASA Infrared Telescope Facility on 2025 September 9 \footnote{The sources were observed as part of Program 2025B103}. The data were reduced and corrected for telluric absorption using standard tools \citep{Vacca2003, Cushing2004}. For total integration times of 16 minutes for WTP15aaavof and 24 minutes for WTP16aakapb, we obtained decent signal-to-noise ratios only in the K-band. The resulting spectra (R$\approx2000$; Figure \ref{fig:yso_irtf}) clearly show CO absorption bands, a signature of FUOrs \citep{Connelley2018} that is not resolved in the low-resolution SPHEREx spectra. We measure equivalent widths (EWs) of the 2.29\,\um~CO band, the Ca~I, and Na~I lines, following \citet{Messineo2021}, and obtain EW(CO) $\approx20$\angstrom, EW(Ca~I + Na~I) $\approx2.0$\angstrom for WTP15aaavof; and EW(CO)$\approx30$\angstrom, EW(Ca I + Na I)$\approx1.5$\angstrom for WTP16aakapb. These values are consistent with those expected for FUOrs, distinguishing them from other young stellar objects \citep{Connelley2018}.  

Based on their locations within star-forming regions, their large amplitude, long-duration mid-IR outbursts, and their SPHEREx and near-IR spectral similarities to the known FUOrs, we suggest that WTP16aakapb and WTP15aaavof are new FUOr sources. We note that further evidence to cement this classification is warranted, such as higher resolution spectra covering shorter wavelengths to search for additional FUOr diagnostics (\citealt{Connelley2018}, Portnoi et al., in prep), and detailed modeling of the outburst and quiescent spectral energy distributions (e.g., \citealt{Frostig2025}), but is beyond the scope of this Letter. Here, we have demonstrated an effective strategy for identifying promising FUOrs, by focusing on large-amplitude, long-duration NEOWISE outbursts whose SPHEREx spectra show cool molecular absorption features. Together, these criteria efficiently exclude common contaminants in searches based on either spectra or lightcurves alone, such as evolved stars that have cool spectra but no large-amplitude eruptions, or classical novae which show large-amplitude outbursts but whose spectra are dominated by strong emission lines (see Section \ref{sec:novae}).

\subsection{Transients with emission lines : Classical novae}
\label{sec:novae}
\begin{figure*}
    \centering
    \includegraphics[width=\textwidth]{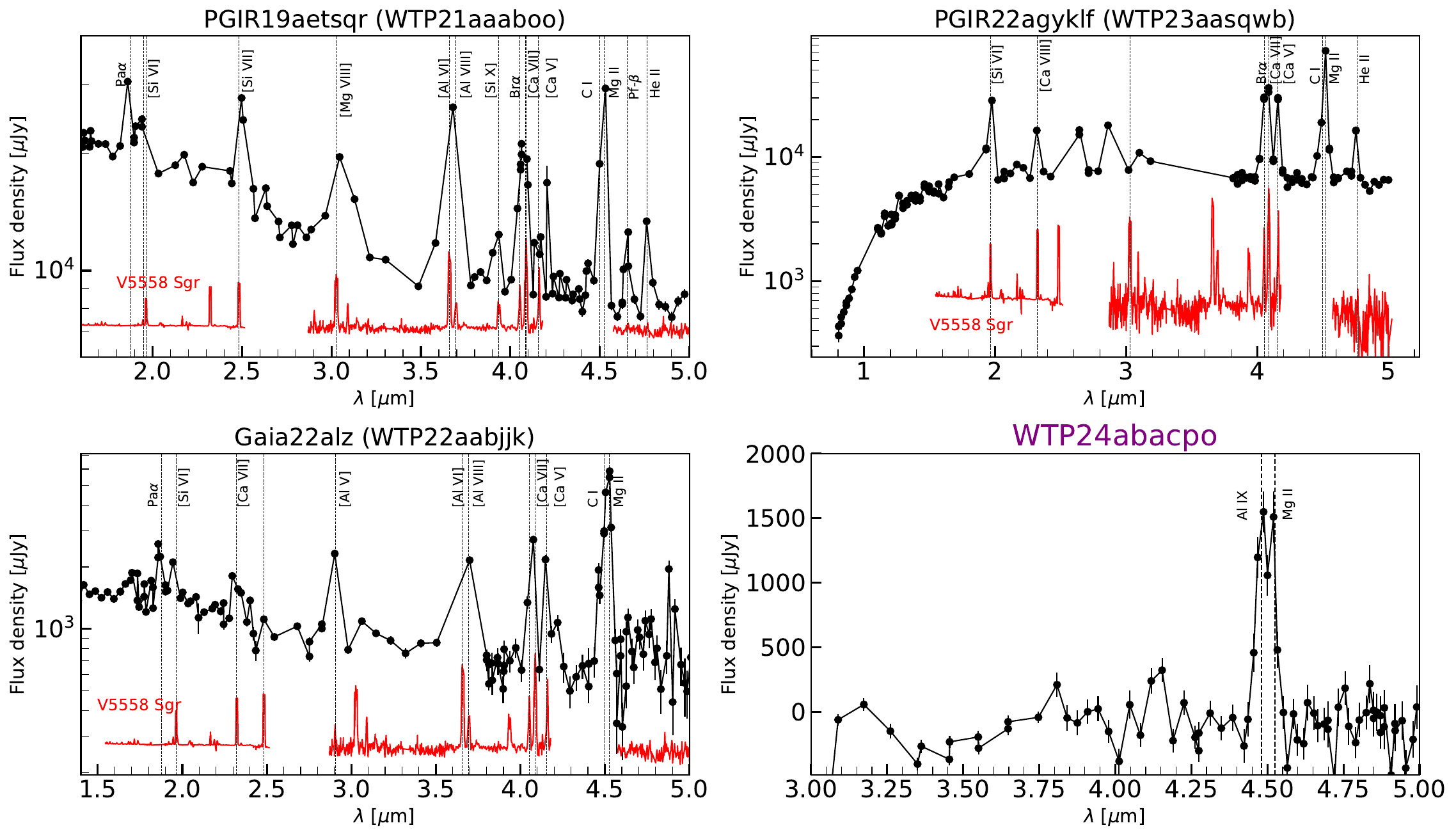}
    \caption{SPHEREx infrared spectra for the four transients showing strong emission lines in their spectra --- three previously known classical novae and one likely new one. Plotted in red, for comparison, are medium resolution IR spectrum of the classical nova V5558 Sgr \citep{Rudy2025a} (scaled and shifted arbitrarily).}
    \label{fig:nova_spectra}
\end{figure*}
Three of our transients are previously known classical novae --- PGIR19aetsqr (WTP21aaaboo, \citealt{De2021}), PGIR22agyklf (WTP23aasqwb, \citealt{De2021}), and Gaia22alz (WTP22aabjjk, \citealt{Aydi2023, 2022ATel15270....1B}). The NEOWISE lightcurves of all three novae show a rapid, large amplitude brightening ($10-100$x increase in flux, Figure \ref{fig:lcs}) reaching peak mid-IR flux $>$10$^{5}$\,$\mu$Jy, followed by a steady brightness decline.  Morphologically, these light curves closely resemble the mid-IR evolution of classical novae previously reported by \citet{Zuckerman2023, Ramesh_2025}. Their SPHEREx spectra (Figure \ref{fig:nova_spectra}) show strong emission lines. We identify emission from hydrogen, carbon, aluminium, and calcium --- consistent with those seen in previous infrared spectra of classical novae \citep{Gehrz1988,Rudy2022, Rudy2025a, Rudy2025b}. This is illustrated by the comparison to a medium resolution IR spectrum of the classical nova V558\,Sgr from \citet{Rudy2025a} shown in Figure \ref{fig:nova_spectra}. Notably, the strongest feature in all three spectra is centered around $\sim$4.5\,\um. This wavelength range is inaccessible from the ground due to atmospheric CO$_2$ absorption, and to our knowledge, this feature has not been previously reported in novae. We suggest that this feature is likely due to one or more of the MgII doublet at 4.5247 and 4.5254\,\um, a CI emission line at 4.49926\,\um, and the Al IX emission line at 4.48019\,\um. 

The newly identified transient WTP24abacpo (RA= 185.7955, Dec= $-61.8313$) was only detected in the final NEOWISE visit to its field in July 2024. It underwent a large amplitude outburst, reaching a brightness similar to the three novae described above (Figure \ref{fig:lcs}). Performing forced photometry on images from the ATlAS survey \citep{Tonry2018}, we recover an optical outburst beginning about sixty days before and lasting until forty days after the NEOWISE epoch. The transient has red colors, with ATLAS $c-o\approx2.5$\,mag and $o-W1\approx3$\,mag, suggesting that it is obscured. The transient is not detected in any ATLAS images taken after November, 2024. In SPHEREx, this source is only detected in the 4-5\,\um~range, and shows the same strong emission feature at $\sim4.5$\,\um~seen in the confirmed novae. Based on its bright, fast-evolving lightcurve and the SPHEREx spectral similarity, we suggest that this transient is a missed Galactic nova\footnote{This source has not been reported to the Transient Name Server, and does not appear in compilations of confirmed Galactic novae : \href{https://asd.gsfc.nasa.gov/Koji.Mukai/novae/novae.html}{https://asd.gsfc.nasa.gov/Koji.Mukai/novae/novae.html}}. 

In addition to these four, we identify many similarly bright NEOWISE outbursts from 2023-2024 that are also likely classical novae, but were not detected by SPHEREx, likely because they faded away before their SPHEREx observations, that are typically obtained more than a year after the last NEOWISE epoch. Nevertheless, our results demonstrate that despite its modest spectral resolution, SPHEREx is capable of detecting strong emission lines in eruptive transients. SPHEREx will thus aid future searches for emission line transients, including classical novae, X-ray binaries \citep{Clark1999}, and symbiotic systems \citep{Merc2025} undergoing outbursts during its mission lifetime. 

\begin{figure}[!hbt]
    \centering
    \includegraphics[width=0.5\textwidth]{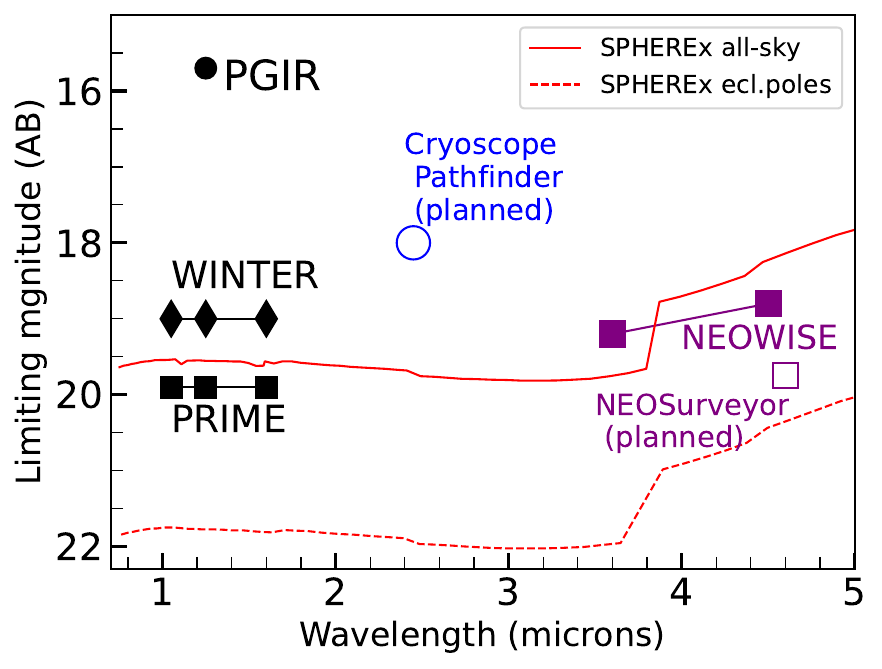}
    \caption{SPHEREx in the context of ongoing and planned IR time-domain surveys. SPHEREx is poised to provide low-resolution spectra for the vast majority of transients identified by these surveys. }
    \label{fig:neowise_histogram}
\end{figure}

\section{Summary and way forward}
\label{sec:summary}
In this Letter, we presented SPHEREx spectra for eight Galactic IR outbursts selected using the NEOWISE survey. We first examined two known FUOrs and three known classical novae, and showed that their low-resolution SPHEREx spectra clearly exhibit hallmarks such as cool molecular absorption in FUOrs and strong emission lines in novae. Using these features, we identified two new likely FUOrs and one missed Galactic nova. Our results highlight the effectiveness of low-resolution SPHEREx spectra in classifying slow IR outbursts. 

The bright examples presented here represent only the tip of the iceberg. Thousands of time-variable mid-IR sources have already been identified in NEOWISE alone (e.g., \citealt{Paz2024}), and mining the NEOWISE datasets together with SPHEREx promises to uncover hidden populations of dusty outbursts. Repeated SPHEREx observations over its two-year mission will further enable the tracing of their spectral evolution. 

We now describe some anticipated challenges. SPHEREx spectroscopy of faint outbursts in crowded Galactic regions is difficult due to SPHEREx's relatively coarse pixel scale, and similar difficulties are expected for extragalactic transients projected against bright host galaxies. In addition, as SPHEREx acquires spectra over a period of $\approx$two weeks or longer, intrinsic variability on comparable or shorter timescales may be imprinted on the resulting spectra for fast-evolving sources. Finally, while SPHEREx will provide initial identifications of interesting outbursts, higher spatial and spectral resolution ground-based spectroscopy will remain essential for detailed characterizations of the most compelling events, which will be particularly challenging for severely obscured events. 

Looking ahead, SPHEREx will provide spectroscopic classifications for transients discovered by other IR surveys, such as the ongoing Palomar Gattini IR, WINTER, and PRIME surveys, and the upcoming Cryoscope survey \citep{Kasliwal2025_cryoscope}. These will set the stage for future, more sensitive IR transient searches with the upcoming NEO Surveyor mission and the \emph{Nancy Grace Roman Space Telescope}. Figure \ref{fig:neowise_histogram} shows SPHEREx in the landscape of these IR time-domain surveys. With its unprecedented, all-sky IR spectroscopic coverage, SPHEREx has opened up a new frontier for IR time-domain astronomy.

\section*{Acknowledgements}
We are immensely grateful to the SPHEREx team for timely release of SPHEREx data and tools that facilitated this work. We also thank David Hogg for useful discussions. VRK was supported by NASA
through the NASA Hubble Fellowship grant \#HST-HF2-51578.001-A awarded by the Space Telescope Science Institute, which is operated by the Association of Universities for Research in Astronomy, Inc., for NASA, under contract NAS5-26555. VRK also acknowledges the hospitality of the CCA-Flatiron Institute. This publication makes use of data products from the Wide-field Infrared Survey Explorer, which is a joint project of the University of California, Los Angeles, and the Jet Propulsion Laboratory/California Institute of Technology, funded
by the National Aeronautics and Space Administration. This publication makes use of data products from the Spectro-Photometer for the History of the Universe, Epoch of Reionization and Ices Explorer (SPHEREx), which is a joint project of the Jet Propulsion Laboratory and the California Institute of Technology, and is funded by the National Aeronautics and Space Administration. We acknowledge the support of the National Aeronautics and Space Administration through ADAP grant number 80NSSC24K0663. The computations reported in this paper were (in part) performed using resources made available by the Flatiron Institute. D.F.'s contribution to this material is based upon work supported by the National Science Foundation under Award No. AST-2401779. J.L.S. acknowledges Columbia Data Science Institute award SF-177. This research award is partially funded by a generous gift of Charles Simonyi to the NSF Division of Astronomical Sciences. The award is made in recognition of significant contributions to Rubin Observatory’s Legacy Survey of Space and Time. The Flatiron Institute is funded by the Simons Foundation. Astronomer observing with the Infrared Telescope Facility, which is operated by the University of Hawaii under contract 80HQTR24DA010 with the National Aeronautics and Space Administration

\section*{Data availability}
This work uses publicly available SPHEREx quick release (QR2) images \citep{SPHEREx_IPAC_dataset}.
The NEOWISE and ATLAS lightcurves, SPHEREx and IRTF spectra described in this paper will be made available on Zenodo upon publication.

\bibliography{myreferences}
\end{document}